\documentclass{INTERSPEECH2023}

\interspeechcameraready 

\usepackage{amsmath,graphicx}
\usepackage[acronym]{glossaries}
\usepackage{url}
\usepackage{amsmath}
\usepackage{pgfplots}
\usepackage{graphicx}
\usepackage{booktabs}
\usepackage{comment}
\usepackage{microtype}
\newcommand\norm[1]{\left\lVert#1\right\rVert}

\definecolor{color1}{HTML}{00b159}
\definecolor{color2}{HTML}{f37735}
\definecolor{color3}{HTML}{00aedb}
\definecolor{color4}{HTML}{ffc425}
\definecolor{color5}{HTML}{d11141}

\title{Controllable Generation of Artificial Speaker Embeddings \\
through Discovery of Principal Directions}

\name{Florian Lux, Pascal Tilli, Sarina Meyer, Ngoc Thang Vu}
\address{University of Stuttgart, Institute for Natural Language Processing, Germany}
\email{florian.lux@ims.uni-stuttgart.de}

\begin{document}

\maketitle
\begin{abstract}
Customizing voice and speaking style in a speech synthesis system with intuitive and fine-grained controls is challenging, given that little data with appropriate labels is available. Furthermore, editing an existing human's voice also comes with ethical concerns. In this paper, we propose a method to generate artificial speaker embeddings that cannot be linked to a real human while offering intuitive and fine-grained control over the voice and speaking style of the embeddings, without requiring any labels for speaker or style. The artificial and controllable embeddings can be fed to a speech synthesis system, conditioned on embeddings of real humans during training, without sacrificing privacy during inference.
\end{abstract}
\section{Introduction and Related Work}
\label{sec:intro}


In recent years, neural text-to-speech (TTS) systems have made significant advances, reaching human levels of naturalness in subjective evaluation \cite{liu2021delightfultts, tan2022naturalspeech} and reaching state-of-the-art results on speaker similarity and quality using just a few minutes of adaptation data \cite{Neekhara2021AdaptingTM, casanova2022yourtts}. Even in the zero-shot adaptation setting, progress is made very quickly. \cite{arik2018neural} first propose using an external speaker encoder as a conditioning signal. \cite{jia2018transfer} pre-train a speaker encoder on a speaker verification task. \cite{cooper2020zero} identify the x-vector embeddings \cite{xvect} which have gained much popularity as a TTS conditioning signal, as insufficient to generalize to unseen voices without loss in quality. Speaker embeddings which select and remix vectors that contain information about the encoded utterance using attention have also been around for a relatively long time \cite{wang2018style, choi2020attentron}, but are still being actively researched \cite{Wu2022AdaSpeech4A}. These models are usually trained jointly with the TTS system, which allows them to encode more prosodic information rather than only a speaker's identity, in contrast to speaker embeddings derived from speaker verification models. Other approaches to encode more information in speaker embeddings include using the voice conversion task, which works especially well together with jointly optimized embedding functions \cite{chien2021investigating}. Further designs that aid the generalization of zero-shot voice cloning in TTS include the use of a reconstruction loss, where the distance between a reference embedding and an embedding of the output speech is minimized \cite{Wu2022AdaSpeech4A, nachmani2018fitting, 9673749}. Also, the hierarchical variational autoencoder (VAE) approach can be used to model characteristics of speakers and speaking styles like a deconstructed speaker representation that can be controlled at multiple levels in the VAE \cite{hsu2018hierarchical}. Although this approach seems very promising in its experimental validation, it lacks reproducibility and is therefore difficult to use and compare with.  


While cloning a person's voice opens the door to many exciting and practical applications, it is problematic from an ethical point of view. A person's voice is closely linked to their identity and personal attributes like age or social background, which can to some extent be detected in recordings of their speech without consent. This leads to serious privacy violations \cite{kroeger2020privacy}. Voice actors are not necessarily aware of the implications and future use of their voice in TTS tools \cite{scott2019who} and might not agree to the content generated with it. Hence it is ethically tricky to use the voice of a real human to synthesize utterances they have never said before. Therefore, voice anonymization approaches using voice conversion techniques attempt to synthesize speech with target voices that are not linked to actual humans. This is usually done by either creating an average voice over a pool of speakers \cite{fang2019speaker}, or by training generative models to sample artificial voices \cite{meyer2022anonymizing,turner2022generating}. Sampling such an artificial voice, however, comes with the problem of having little to no control over how the resulting voice will sound like, or the voice distributions will sound unnatural \cite{fang2019speaker, turner2022generating}. Recently, \cite{van2022voiceme} proposed to design a speaker embedding by modifying the principal components of an already existing one. While this enables controllability, it requires a real human's embedding, which again comes with the possibility of misuse. A further challenge is that while controllability is a highly sought-after property in both TTS and voice privacy, it is notoriously hard to quantify, which in most cases completely rules out comparisons between approaches. This means that any new proposed approach can only be evaluated to the extent of whether it works itself. A lack of open-source codebases or models for such controllability approaches that operate on a high level intensifies the complications of comparing approaches. 


To remedy the problems of speaker embeddings being either not controllable, not natural, or not useful for privacy applications, we propose to train a Wasserstein Generative Adversarial Network (WGAN) \cite{gan_goodfellow, wgan, improved_wgan} as we previously described in \cite{meyer2022anonymizing} that is capable of generating new points in the speaker embedding space which are not associated to any real human. This process is then made controllable by discovering principal directions in the latent space of the GAN, as proposed in the GANSpace approach for the image domain \cite{harkonen2020ganspace}. The generated embeddings are inspected using an auxiliary TTS system to explore the nature of the changes applied using the controls. With this approach, we can generate natural-like artificial speaker embeddings that match desired properties without any way of tracing the resulting speaker embedding back to a human speaker and without the need for any labeled data regarding the features that we want to control. We evaluate the controllability and privacy aspects of our proposed approach in Section \ref{sec:exp:control} by using objective measures to quantify the impact of the control mechanisms on the output. The final interface to the embedding modification can be realized as a set of sliders that can be used to control intuitive properties of the embedding without any expert knowledge. 


Summarizing our contributions: 1) we propose a framework to modify intuitively understandable properties of speaker embeddings, and 2) we do so in a setup that does not use references of actual humans as the basis. We verify our contributions in experiments using objective evaluation and provide all code and models, as well as an interactive demo, open-source\footnote{\url{https://github.com/DigitalPhonetics/IMS-Toucan}}.

\section{Proposed Method}

\begin{figure}[t]
    \centering
    \includegraphics[width=\columnwidth]{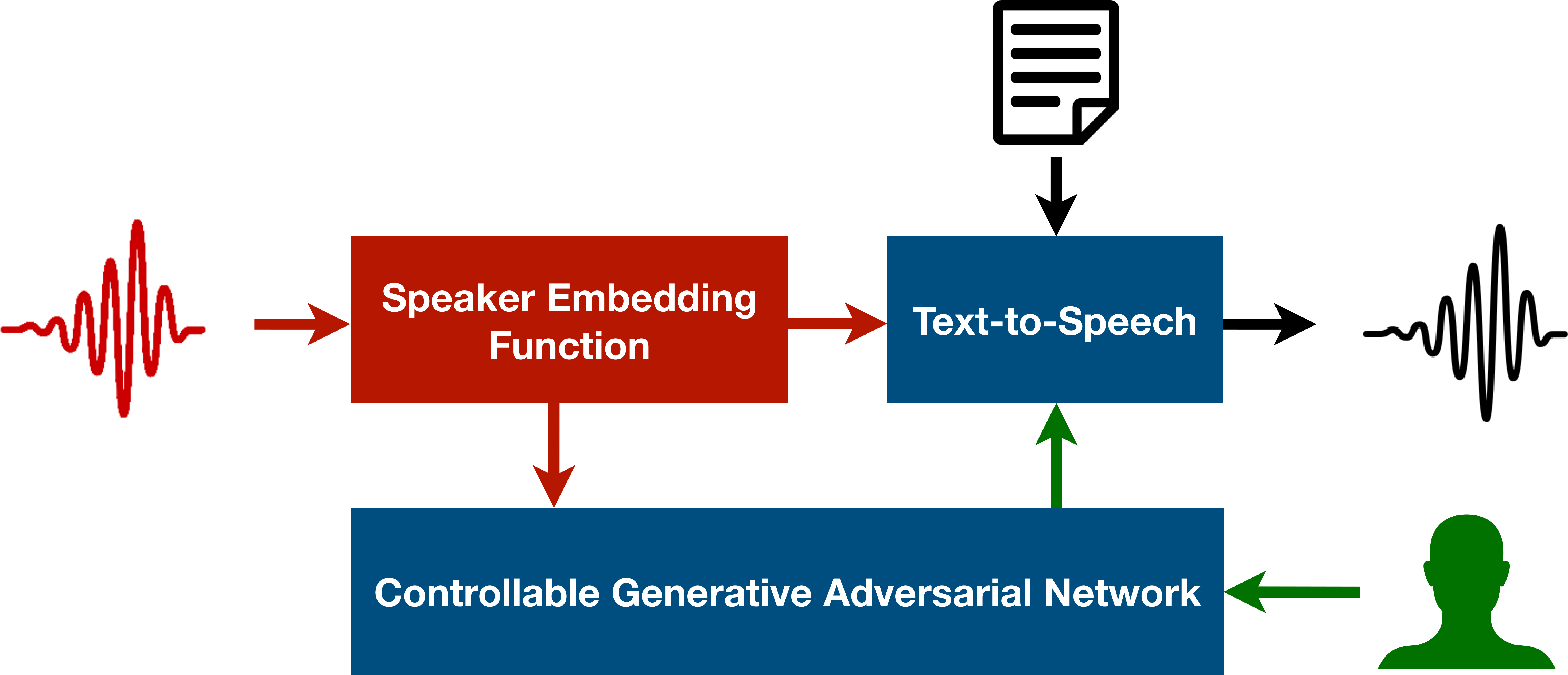}
    \caption{Overview of the proposed approach: The red arrows exist only during training time. During inference, only the green arrows exist. Hence, during inference, there is no link to an input speech signal, that could be traced back to a real human. The embeddings that the GAN produces can be controlled during inference, but require no labels during training.}
    \label{fig:overview}
\end{figure}

\subsection{Overview}
Figure \ref{fig:overview} shows an overview of our proposed method. We train a speech embedding function following the Global Style Token (GST) approach \cite{wang2018style} jointly with a TTS system on monolingual expressive multispeaker data. We then use the speaker embedding function in a frozen state as a conditioning signal to another multilingual and multispeaker TTS system, outlined in Section \ref{sec:setup}. We use the embedding function to sample many unsupervised speaker embeddings from various multispeaker datasets further described in Section \ref{sec:data}. These sampled speaker embeddings are used as the target distribution of a WGAN model, which learns to generate novel points in the distribution of diverse speaker embeddings, explained in Section \ref{sec:generation}. Training the WGAN requires no labels for style and speaker, however by adding a control mechanism on top of the WGAN, we can find intuitive controls over the generated speaker embeddings that can be aligned with categorical labels through empirical exploration, as described in Section \ref{sec:control}. This can be done during inference time, where the speaker embedding function is no longer used, and therefore, all ties to an input spectrogram are cut such that no speaker's privacy is violated. The speaker embeddings used during inference time are purely synthetic and can be controlled intuitively without expert knowledge.

\subsection{Controllable Generative Adversarial Network}
\subsubsection{Speaker Embedding Generation}
\label{sec:generation}
We rely on a WGAN with Quadratic Transport Cost, as proposed in \cite{meyer2022anonymizing} 
to generate artificial speaker embeddings. Our WGAN learns to map random noise vectors into the distribution of speaker embeddings derived from human speech on an utterance level. Similar to the initial GAN approach proposed by \cite{gan_goodfellow}, the generator receives an input vector $z$ randomly sampled from a normal distribution $\mathcal{N}(0,1)$. Since we are using a WGAN, the objective function is changed to computing the Wasserstein distance between real and artificial data, as introduced by \cite{wgan, improved_wgan}. To improve the convergence properties of WGAN, we follow \cite{wgan_qc} in additionally computing the quadratic transport cost.

We sample a $z$, generate an embedding from it and then directly feed it into our TTS model to elicit a voice and speaking style. The probability of sampling a new speaker embedding vector in a continuous 64 dimensional space indistinguishable from a speaker embedding extracted from an actual human is not zero but highly unlikely. Consequently, we can sample voices that do not exist.

\subsubsection{Controlling the Generation Process}
\label{sec:control}
Since our objective is to control the voice and speaking style of our TTS system, sampling random speaker embeddings from the WGAN is not sufficient. We need to intervene in the generation process, which is random due to the input of the GAN being randomly sampled vectors from $\mathcal{N}(0,1)$. Searching directly in the prior distribution $p(z)$ for directions does not help us since the distribution is isotropic.
However, \cite{harkonen2020ganspace} found that principal components of feature tensors within the early layers of GANs contain the most information and variation. We leverage this property by computing the principal components using Principal Component Analysis (PCA). First, we sample $N$ latent vectors $z_{1:N}$. Afterward, we propagate these vectors through our generator and save the intermediate feature representations $y_{1:N}$, which are computed by the first layer of the GAN. We then use $y_{1:N}$ to compute a low-rank basis matrix $V$ with PCA, as well as the mean $\mu$ of $y_{1:N}$. The PCA coordinates $X$ are computed by $X=(Y - \mu) V$, where $Y$ is the matrix containing the feature representations $y_{1:N}$. Finally, we compute a basis $U$ as shown in Equation \ref{eq:u} using a least-squares solver.
\begin{equation}
    U = \text{arg min} \sum_{j} \norm{U x_j - z_j}^2
    \label{eq:u}
\end{equation}
The columns $u_k$ of $U$ are called principal directions. Now we are able to modify the latent vector $z$ along the principal directions with $z^{\prime} = z + Ux$, where $x$ is a vector that specifies the offsets of each column $u_k$. To determine which principal direction corresponds to which surface level property, we use an auxiliary TTS system that is conditioned on the frozen embedding function used to train the GAN. We input embeddings, modify them and observe what changes in the synthesized speech.

\section{Experimental Setup}

\subsection{Model Configuration}
\label{sec:setup}
We use FastSpeech 2 \cite{ren2020fastspeech} as the synthesis architecture with phoneme averaged pitch and energy, as suggested in FastPitch \cite{lancucki2021fastpitch}, which enables a significant amount of fine-grained control over the produced speech already. We use articulatory features as the input, as suggested in \cite{lux2022laml}. We also use a flow-based PostNet, as suggested in PortaSpeech \cite{ren2021portaspeech}. To perform spectrogram inversion, we use the Avocodo architecture \cite{bak2022avocodo}, which is based on HiFiGAN \cite{kong2020hifi}. The embedding function used is the attentive and jointly trained GST approach \cite{wang2018style}, and the embedding GAN consists of small ResNet blocks \cite{he2016deep} in both the generator and discriminator. We implement all of this in our open source toolkit, IMS Toucan \cite{lux2021toucan}.

\begin{figure}[t]
    \centering
    \includegraphics[trim={3,5cm, 17.5cm, 3cm, 2.3cm}, clip, width=.7\columnwidth]{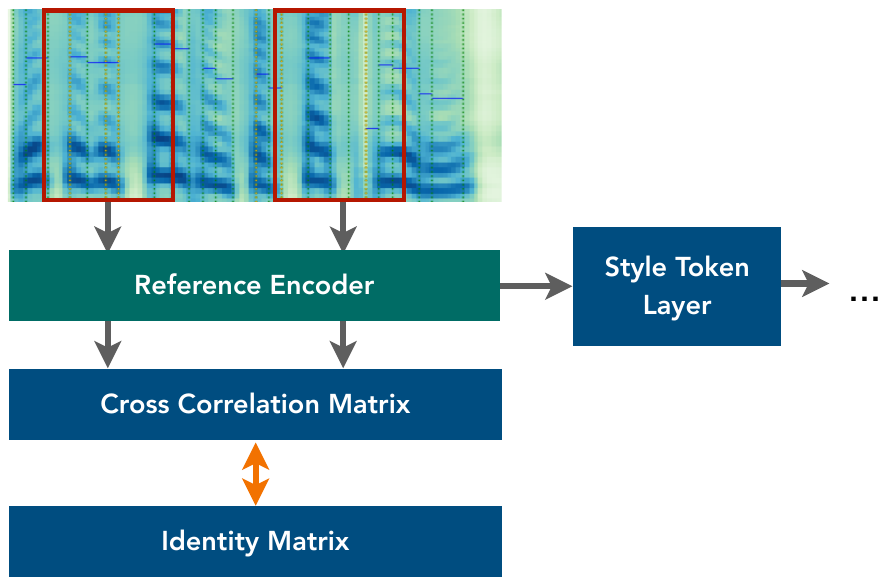}
    \caption{Visualization of the Barlow Twins loss for the speaker embeddings. The red windows are taken at random positions from the spectrogram and embedded. The orange arrow depicts an $L_1$ distance, that is minimized as additional objective.}
    \label{fig:bt_emb_loss}
\end{figure}

Further, we add a mechanism that helps the joint training of embedding function and TTS converge with around half the number of steps it usually takes: We use the Barlow Twins objective \cite{zbontar2021barlow} on the intermediate representation before the attentive layers in the GST embedding function. To sample the positive pair needed for this objective, we take two random windows from the same signal and assume that the signal is consistent in style. This is visualized in Figure \ref{fig:bt_emb_loss}. This redundancy reduction objective rewards preserving more information of the reference signal, which the attentive layers can then pick up on. 

\subsection{Data Used}
\label{sec:data}
We use a combination of three datasets to train the embedding function, the TTS and the GAN.
The first is LibriTTS \cite{zen2019libritts}, a large-scale English multispeaker dataset comprising 585 hours of audio books read by a total of 2,456 speakers. We include two additional datasets with acted emotional speech to add more versatile data. However, we only utilize their speech and text components without any emotion labels. 
The first is the Ryerson Audio-Visual Database of Emotional Speech and Song (RAVDESS) \cite{livingstone2018ryerson}. It consists of 7,356 audio samples performed by 24 speakers. The second is the Emotional Speech Dataset (ESD) \cite{zhou2021emotional}. 
It consists of English and Mandarin utterances but we only use the 1,750 samples by 10 English speakers. 
Once the embedding function is trained jointly with a monolingual English TTS model, we freeze its parameters and use it as a conditioning signal to train a multilingual TTS model, as described in \cite{lux2022low}. This multilingual TTS model is trained on a total of 14 languages to showcase that our proposed approach can be easily applied to more than just English benchmark data and thus bring the advances of controllable voice privacy systems to a decent portion of the world's population. The data we used to train this multilingual TTS system is shown in table \ref{tab:data}.

\begin{table}[ht]
    \centering
    \begin{tabular}{l|l|c}
    \toprule
        Dataset & Language & Hours Used \\ \midrule
        Blizzard Challenge 2011 \cite{king2011blizzard}                    & English     & 5 \\
        LJSpeech  \cite{ljspeech17}                                        & English     & 10 \\
        LibriTTS \cite{zen2019libritts}                                    & English     & 50 \\
        HiFi-TTS \cite{bakhturina2021hi}                                   & English     & 20 \\
        VCTK \cite{veaux2017superseded}                                    & English     & 5 \\
        HUI-Audio-Corpus \cite{puchtler2021huiaudiocorpusgerman}           & German      & 60 \\
        Thorsten \cite{muller_thorsten_2021_5525342}                       & German      & 10 \\
        Blizzard Challenge 2021 \cite{ling2021blizzard}                    & Spanish     & 5 \\
        CSS10 \cite{css10}                                                 & Spanish     & 20 \\
        CSS10 \cite{css10}                                                 & Greek       & 4 \\
        CSS10 \cite{css10}                                                 & Finnish     & 11 \\
        CSS10 \cite{css10}                                                 & French      & 5 \\
        Multilingual LibriSpeech \cite{pratap2020mls}                      & French      & 34 \\
        CSS10 \cite{css10}                                                 & Russian     & 21 \\
        CSS10 \cite{css10}                                                 & Hungarian   & 10 \\
        CSS10 \cite{css10}                                                 & Dutch       & 5 \\
        Multilingual LibriSpeech \cite{pratap2020mls}                      & Dutch       & 29 \\
        Multilingual LibriSpeech \cite{pratap2020mls}                      & Polish      & 20 \\
        Multilingual LibriSpeech \cite{pratap2020mls}                      & Portuguese  & 25 \\
        Multilingual LibriSpeech \cite{pratap2020mls}                      & Italian     & 30 \\
        Aishell-3 \cite{shi2020aishell}                                    & Chinese     & 70 \\
        InfoRe Technology 1 \cite{infore}                                  & Vietnamese  & 10 \\
        \bottomrule
    \end{tabular}
    \caption{Data across languages used to train the multilingual TTS model, to which we apply our proposed control mechanism.}
    \label{tab:data}
\end{table}

\subsection{Experiments}

\subsubsection{Controllability}
Since it is challenging to quantify controllability, we measure the effect of controlling a property in a speaker embedding on the output of the TTS system that takes the embedding as a condition. Even this is challenging for high-level properties, such as timbre, sibilance, microphone characteristics or room acoustics. Therefore, we select two properties of the speech for which we can confidently provide reliable objective measures. We use auxiliary trained classifiers to evaluate to which extent gender (on a continuous scale from masculine to feminine) and arousal (on a scale from low to high) can be modified using the principal directions. We select the two corresponding axes in the latent space and sample 300 GAN-generated artificial speaker embeddings for this. For each embedding and property, we move the latents in consistent steps along the corresponding directions and synthesize three utterances consisting of each six phonetically balanced sentences taken from \cite{rothauser1969ieee}. This results in 18,000 audios per property and 60 per sampled speaker embedding. For each audio, we apply high-performing open-source models to predict gender\footnote{\url{https://huggingface.co/versae/wav2vec2-base-finetuned-coscan-sex}} and arousal\footnote{\url{https://huggingface.co/audeering/wav2vec2-large-robust-12-ft-emotion-msp-dim}} of these entirely artificial voices. We verify that both of these models reach state-of-the-art on common benchmark datasets for their respective tasks to ensure their predictions are reliable.

\subsubsection{Privacy}
To investigate whether neither the TTS, the embedding function, nor the GAN gravitates towards speakers seen in training, we synthesize speech of 1,000 artificial speaker embeddings and use two separate speaker verification models to verify that the artificial speakers are never verified as any speaker from the training set such that they can be considered novel non-existing speakers.
The speaker verification models are provided off-the-shelf by the SpeechBrain toolkit \cite{speechbrain} and follow the x-vector \cite{xvect} and ECAPA-TDNN \cite{ecapa} architectures.

\section{Results}

\subsection{Controlling Gender}
\label{sec:exp:control}

\begin{figure}[ht]
    \centering
    \begin{tikzpicture}
        \begin{axis}[%
        width=0.95\columnwidth,
        height=0.5\columnwidth,
        xmin=-50.02, xmax=50.02,
        ymin=0,
        ylabel={samples in \%},
        xlabel={points along principal direction},
        ]
        \addplot[ybar stacked, bar width=8, xshift=0.5*\pgfplotbarwidth, color=color2, fill=color2]%
        coordinates{
        (-25.0,0.0)(-20.0,0.0)(-15.0,0.0)(-10.0,1.33)(-5.0,10.0)(0.0,20.0)(5.0,15.0)(10.0,7.0)(15.0,0.33)(20.0,0.0)
        };
        
        \addplot[ybar stacked, bar width=8, xshift=0.5*\pgfplotbarwidth, color=color1, fill=color1]%
        coordinates{
        (-25.0,0.0)(-20.0,0.0)(-15.0,0.0)(-10.0,2.33)(-5.0,6.67)(0.0,16.33)(5.0,15.67)(10.0,5.33)(15.0,0.0)(20.0,0.0)
        };
        \legend{F $\rightarrow$ M, M $\rightarrow$ F}
        \end{axis}
    \end{tikzpicture}
    \caption{Flip points of prediction change on the \textbf{F}eminine/\textbf{M}asculine scale.}
    \label{fig:gender}
\end{figure}
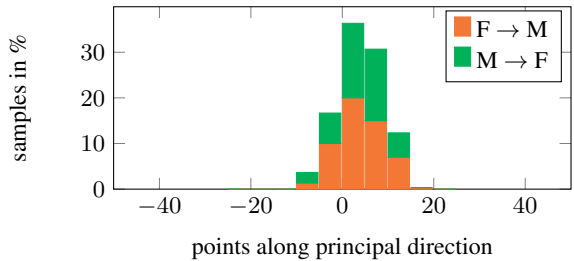

For the gender prediction, we observe either feminine or masculine very clearly, without any predictions being uncertain. The prediction switches for all voices we tested at some point as we move along the axis. However, the direction of movement towards turning the prediction into feminine or masculine is not the same for each voice, nor is the prediction the same for each sampled base embedding (54\% switch F$\rightarrow$M, the remaining 46\% switch M$\rightarrow$F). These flipping points and corresponding numbers of voices are given in Figure \ref{fig:gender}. All prediction switches occur around the center of the axis, between values of -10 and 15. Overall, the experiment shows that the given principal direction clearly affects features in the speaker embeddings leading to the perception and prediction of the voice being either masculine or feminine.

\subsection{Controlling Arousal}

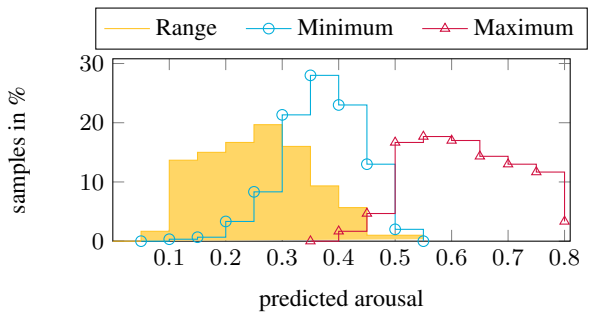
\begin{figure}[ht]
    \centering
    \begin{tikzpicture}
        \begin{axis}[%
        width=0.95\columnwidth,
        height=0.5\columnwidth,
        xmin=0.0, xmax=0.81,
        ymin=0,
        legend columns=3,
        legend style={at={(0.5,1.05)},anchor=south,
            /tikz/every even column/.append style={column sep=0.3cm}},
        ylabel={samples in \%},
        xlabel={predicted arousal},
        xtick={0.1,0.2,0.3,0.4,0.5,0.6,0.7,0.8}
        ]
        \addplot[const plot, color=color4, fill=color4!70]%
        coordinates{
       (0.0,0.0)(0.05,1.67)(0.1,13.67)(0.15,15.0)(0.2,16.67)(0.25,19.67)(0.3,16.0)(0.35,9.33)(0.4,5.67)(0.45,1.0)(0.5,1.0)(0.55,0.33)
        };
        \addplot[const plot, color=color3, mark=o]%
        coordinates{
        (0.05,0.0)(0.1,0.33)(0.15,0.67)(0.2,3.33)(0.25,8.33)(0.3,21.33)(0.35,28.0)(0.4,23.0)(0.45,13.0)(0.5,2.0)(0.55,0.0)
        };
        \addplot[const plot, color=color5, mark=triangle]%
        coordinates{
        (0.35,0.0)(0.4,1.67)(0.45,4.67)(0.5,16.67)(0.55,17.67)(0.6,17.0)(0.65,14.33)(0.7,13.0)(0.75,11.67)(0.8,3.33)
        };
        \legend{Range, Minimum, Maximum}
        \end{axis}  
    \end{tikzpicture}
    \caption{Minimum, maximum, and their difference (range) of predicted arousal across the principal direction.}
    \label{fig:arousal}
\end{figure}

For arousal, the results are less apparent than gender because the classifier predicts a continuous value between 0.0 and 1.0 instead of binary results. However, movement along the respective axis affects the prediction. It is visualized in Figure \ref{fig:arousal} using three metrics: the minimal predicted arousal value for each seed speaker, the maximal one, and the distance between minimum and maximum (range). For each metric, the relative amount of seed speakers is given to the corresponding arousal values, grouped in bins of 0.05 values. For instance, 28\% of generated speaker embeddings can be modified to a predicted arousal of between 0.35 and 0.4 by changing the respective latents. The experiment shows that the arousal of most speaker embeddings can be easily modified in this way. However, the range and level of arousal differ depending on the embedding.


\subsection{Privacy}

Table \ref{tab:er} shows the error rate (ER) of the two speaker verification models when comparing artificial data to train data. Ideally, we would like to see a value of 0\% errors (an error, in this case, meaning that the model verified two samples as being the same speaker while we assume that they should not be). However, since the equal error rate for the two models reported by SpeechBrain is at 3.2\% and 0.8\% respectively, for their test splits, the deviation from 0 is likely due to false positives. Given these results, we conclude that the speech our proposed approach produces is not linked to an actual human.

\begin{table}[ht]
    \centering
    \begin{tabular}{lc}
    \toprule
         & ER \\ \midrule
        X-Vector &  2.1\% \\
        ECAPA-TDNN &  0.9\% \\
        \bottomrule
    \end{tabular}
    \caption{Error rate (ER) of speaker verification systems comparing artificial speakers to all utterances in the train data.}
    \label{tab:er}
\end{table}

\section{Discussion}
\label{sec:discussion}
Whenever we sample a new latent, we generate a point in the speaker embedding space, which we can move around using the principal direction modification. However, the high-dimensional nature of the embedding space makes it near impossible to pinpoint an exact target location and modify the latents accordingly. So if one was trying to design a voice to imitate a real person, e.g., to impersonate them with control over properties like their level of arousal, they could quickly get to a voice with similar properties, however almost certainly not to one that is exactly the same. Hence, our approach allows nuanced control only over artificially generated voices. It is valuable for the future of customizable voice interfaces to limit misuse. We also see a major application of this approach in the field of voice privacy, such that humans can hide the identity of their voice without using another person's voice as mask while being able to control how their masked voice should sound. Also, even though Greek, Finnish, Russian, and Hungarian each only have a single speaker in our train data, we can generate an infinite amount of controllable artificial speakers in those languages, which opens the door to voice privacy applications in plenty of languages.

\section{Conclusion}
In this paper, we present an approach for controlling intuitive properties of artificial speaker embeddings. The embeddings can be used in voice privacy applications and are controlled by modifying intermediate results in the generative network that produces them. We show the effectiveness of this method using objective evaluation by measuring the impact of the control axes on the generated output.

\bibliographystyle{IEEEtran}
\bibliography{main}

\begin{thebibliography}{10}
\providecommand{\url}[1]{#1}
\csname url@samestyle\endcsname
\providecommand{\newblock}{\relax}
\providecommand{\bibinfo}[2]{#2}
\providecommand{\BIBentrySTDinterwordspacing}{\spaceskip=0pt\relax}
\providecommand{\BIBentryALTinterwordstretchfactor}{4}
\providecommand{\BIBentryALTinterwordspacing}{\spaceskip=\fontdimen2\font plus
\BIBentryALTinterwordstretchfactor\fontdimen3\font minus
  \fontdimen4\font\relax}
\providecommand{\BIBforeignlanguage}[2]{{%
\expandafter\ifx\csname l@#1\endcsname\relax
\typeout{** WARNING: IEEEtran.bst: No hyphenation pattern has been}%
\typeout{** loaded for the language `#1'. Using the pattern for}%
\typeout{** the default language instead.}%
\else
\language=\csname l@#1\endcsname
\fi
#2}}
\providecommand{\BIBdecl}{\relax}
\BIBdecl

\bibitem{liu2021delightfultts}
Y.~Liu, Z.~Xu, G.~Wang, K.~Chen, B.~Li \emph{et~al.}, ``{DelightfulTTS: The
  Microsoft Speech Synthesis System for Blizzard Challenge 2021},''
  \emph{{Proc. Blizzard Challenge Workshop}}, 2021.

\bibitem{tan2022naturalspeech}
X.~Tan \emph{et~al.}, ``Naturalspeech: End-to-end text to speech synthesis with
  human-level quality,'' \emph{arXiv:2205.04421}, 2022.

\bibitem{Neekhara2021AdaptingTM}
P.~Neekhara, J.~Li, and B.~Ginsburg, ``{Adapting TTS models For New Speakers
  using Transfer Learning},'' \emph{{arXiv:2110.05798}}, 2021.

\bibitem{casanova2022yourtts}
E.~Casanova, J.~Weber, C.~D. Shulby, A.~C. Junior, E.~G{\"o}lge, and M.~A.
  Ponti, ``{YourTTS: Towards zero-shot multi-speaker tts and zero-shot voice
  conversion for everyone},'' in \emph{{ICML}}, 2022.

\bibitem{arik2018neural}
S.~Arik, J.~Chen, K.~Peng, W.~Ping, and Y.~Zhou, ``{Neural voice cloning with a
  few samples},'' \emph{{NeurIPS}}, 2018.

\bibitem{jia2018transfer}
Y.~Jia, Y.~Zhang, R.~Weiss, Q.~Wang, J.~Shen \emph{et~al.}, ``{Transfer
  Learning from Speaker Verification to Multispeaker Text-To-Speech
  Synthesis},'' in \emph{{NeurIPS}}, 2018.

\bibitem{cooper2020zero}
E.~Cooper, C.-I. Lai, Y.~Yasuda, F.~Fang, X.~Wang \emph{et~al.}, ``{Zero-shot
  multi-speaker text-to-speech with state-of-the-art neural speaker
  embeddings},'' in \emph{{ICASSP}}, 2020.

\bibitem{xvect}
D.~Snyder, D.~Garcia-Romero, A.~McCree, G.~Sell, D.~Povey, and S.~Khudanpur,
  ``{Spoken Language Recognition using X-vectors},'' in \emph{Odyssey 2018},
  2018.

\bibitem{wang2018style}
Y.~Wang, D.~Stanton, Y.~Zhang, R.-S. Ryan, E.~Battenberg \emph{et~al.},
  ``{Style tokens: Unsupervised style modeling, control and transfer in
  end-to-end speech synthesis},'' in \emph{{ICML}}, 2018.

\bibitem{choi2020attentron}
S.~Choi, S.~Han, D.~Kim, and S.~Ha, ``{Attentron: Few-Shot Text-to-Speech
  Utilizing Attention-Based Variable-Length Embedding},'' \emph{Interspeech},
  2020.

\bibitem{Wu2022AdaSpeech4A}
Y.~Wu, X.~Tan, B.~Li, L.~He, S.~Zhao, R.~Song, T.~Qin, and T.-Y. Liu,
  ``{AdaSpeech 4: Adaptive Text to Speech in Zero-Shot Scenarios},'' in
  \emph{Proc. Interspeech 2022}, 2022.

\bibitem{chien2021investigating}
C.-M. Chien, J.-H. Lin, C.-y. Huang, P.-c. Hsu, and H.-y. Lee, ``Investigating
  on incorporating pretrained and learnable speaker representations for
  multi-speaker multi-style text-to-speech,'' in \emph{ICASSP}, 2021.

\bibitem{nachmani2018fitting}
E.~Nachmani, A.~Polyak, Y.~Taigman, and L.~Wolf, ``{Fitting new speakers based
  on a short untranscribed sample},'' in \emph{ICML}, 2018.

\bibitem{9673749}
K.~Azizah and W.~Jatmiko, ``{Transfer Learning, Style Control, and Speaker
  Reconstruction Loss for Zero-Shot Multilingual Multi-Speaker Text-to-Speech
  on Low-Resource Languages},'' \emph{{IEEE Access}}, 2022.

\bibitem{hsu2018hierarchical}
W.-N. Hsu, Y.~Zhang, R.~J. Weiss, H.~Zen, Y.~Wu, Y.~Wang, Y.~Cao, Y.~Jia,
  Z.~Chen, J.~Shen \emph{et~al.}, ``Hierarchical generative modeling for
  controllable speech synthesis,'' in \emph{ICLR}, 2018.

\bibitem{kroeger2020privacy}
J.~L. Kr{\"o}ger, O.~H.-M. Lutz, and P.~Raschke, ``Privacy implications of
  voice and speech analysis -- information disclosure by inference,''
  \emph{Privacy and Identity Management. Data for Better Living: AI and
  Privacy}, 2020.

\bibitem{scott2019who}
K.~M. Scott, S.~Ashby, D.~A. Braude, and M.~P. Aylett, ``Who owns your voice?
  ethically sourced voices for non-commercial tts applications,'' in
  \emph{Proc. International Conference on Conversational User Interfaces},
  2019.

\bibitem{fang2019speaker}
F.~Fang, X.~Wang, J.~Yamagishi, I.~Echizen, M.~Todisco, N.~Evans, and J.-F.
  Bonastre, ``{Speaker Anonymization Using X-vector and Neural Waveform
  Models},'' in \emph{Proc. 10th ISCA Speech Synthesis Workshop}, 2019, pp.
  155--160.

\bibitem{meyer2022anonymizing}
S.~Meyer, P.~Tilli, P.~Denisov, F.~Lux, J.~Koch, and N.~T. Vu, ``Anonymizing
  speech with generative adversarial networks to preserve speaker privacy,'' in
  \emph{{Proc. IEEE SLT}}, 2023.

\bibitem{turner2022generating}
H.~Turner, G.~Lovisotto, and I.~Martinovic, ``Generating identities with
  mixture models for speaker anonymization,'' \emph{Computer Speech \&
  Language}, 2022.

\bibitem{van2022voiceme}
P.~{van Rijn}, S.~Mertes, D.~Schiller, P.~Dura, H.~Siuzdak, P.~M.~C. Harrison,
  E.~André, and N.~Jacoby, ``{VoiceMe: Personalized voice generation in
  TTS},'' in \emph{Proc. Interspeech 2022}, 2022.

\bibitem{gan_goodfellow}
I.~Goodfellow, J.~Pouget-Abadie, M.~Mirza, B.~Xu, D.~Warde-Farley, S.~Ozair,
  A.~Courville, and Y.~Bengio, ``{Generative adversarial nets},''
  \emph{NeurIPS}, 2014.

\bibitem{wgan}
M.~Arjovsky, S.~Chintala, and L.~Bottou, ``{Wasserstein generative adversarial
  networks},'' in \emph{ICML}, 2017.

\bibitem{improved_wgan}
I.~Gulrajani, F.~Ahmed, M.~Arjovsky, V.~Dumoulin, and A.~C. Courville,
  ``{Improved training of Wasserstein GANs},'' \emph{NeurIPS}, 2017.

\bibitem{harkonen2020ganspace}
E.~H{\"a}rk{\"o}nen, A.~Hertzmann, J.~Lehtinen, and S.~Paris, ``{GANSpace:
  Discovering interpretable GAN controls},'' \emph{NeurIPS}, 2020.

\bibitem{wgan_qc}
H.~Liu, X.~Gu, and D.~Samaras, ``{Wasserstein GAN with quadratic transport
  cost},'' in \emph{Proc. ICCV}, 2019.

\bibitem{ren2020fastspeech}
Y.~Ren, C.~Hu, X.~Tan, T.~Qin, S.~Zhao \emph{et~al.}, ``{FastSpeech 2: Fast and
  High-Quality End-to-End Text to Speech},'' in \emph{{ICLR}}, 2020.

\bibitem{lancucki2021fastpitch}
A.~{\L}a{\'n}cucki, ``{FastPitch: Parallel text-to-speech with pitch
  prediction},'' in \emph{{ICASSP}}, 2021.

\bibitem{lux2022laml}
F.~Lux and N.~T. Vu, ``{Language-Agnostic Meta-Learning for Low-Resource
  Text-to-Speech with Articulatory Features},'' in \emph{{ACL}}, 2022.

\bibitem{ren2021portaspeech}
Y.~Ren, J.~Liu, and Z.~Zhao, ``{Portaspeech: Portable and high-quality
  generative text-to-speech},'' \emph{NeurIPS}, 2021.

\bibitem{bak2022avocodo}
T.~Bak, J.~Lee, H.~Bae, J.~Yang, J.-S. Bae, and Y.-S. Joo, ``Avocodo:
  Generative adversarial network for artifact-free vocoder,''
  \emph{arXiv:2206.13404}, 2022.

\bibitem{kong2020hifi}
J.~Kong, J.~Kim, and J.~Bae, ``{HiFi-GAN: Generative Adversarial Networks for
  Efficient and High Fidelity Speech Synthesis},'' \emph{{NeurIPS}}, 2020.

\bibitem{he2016deep}
K.~He, X.~Zhang, S.~Ren, and J.~Sun, ``Deep residual learning for image
  recognition,'' in \emph{CVPR}, 2016.

\bibitem{lux2021toucan}
F.~Lux, J.~Koch, A.~Schweitzer, and N.~T. Vu, ``{The IMS Toucan system for the
  Blizzard Challenge 2021},'' in \emph{{Proc. Blizzard Challenge Workshop}},
  2021.

\bibitem{zbontar2021barlow}
J.~Zbontar, L.~Jing, I.~Misra, Y.~LeCun, and S.~Deny, ``{Barlow twins:
  Self-supervised learning via redundancy reduction},'' in \emph{{ICML}}, 2021.

\bibitem{zen2019libritts}
H.~Zen, V.~Dang, R.~Clark, Y.~Zhang, R.~J. Weiss \emph{et~al.}, ``{LibriTTS: A
  Corpus Derived from LibriSpeech for Text-to-Speech},'' in
  \emph{{Interspeech}}, 2019.

\bibitem{livingstone2018ryerson}
S.~R. Livingstone and F.~A. Russo, ``{The Ryerson Audio-Visual Database of
  Emotional Speech and Song (RAVDESS)},'' \emph{PloS one}, 2018.

\bibitem{zhou2021emotional}
K.~Zhou, B.~Sisman, R.~Liu, and H.~Li, ``{Emotional voice conversion: Theory,
  databases and ESD},'' \emph{Speech Communication}, 2022.

\bibitem{lux2022low}
F.~Lux, J.~Koch, and N.~T. Vu, ``{Low-Resource Multilingual and Zero-Shot
  Multispeaker TTS},'' in \emph{Proc. AACL}.\hskip 1em plus 0.5em minus
  0.4em\relax Association for Computational Linguistics, 2022, pp. 741--751.

\bibitem{king2011blizzard}
S.~King and V.~Karaiskos, ``{The Blizzard Challenge 2011},'' in \emph{{Proc.
  Blizzard Challenge Workshop}}, 2011.

\bibitem{ljspeech17}
K.~Ito and L.~Johnson, ``{The LJ Speech Dataset},''
  \url{https://keithito.com/LJ-Speech-Dataset/}, 2017.

\bibitem{bakhturina2021hi}
E.~Bakhturina, V.~Lavrukhin, B.~Ginsburg, and Y.~Zhang, ``{Hi-Fi Multi-Speaker
  English TTS Dataset},'' in \emph{{Interspeech}}, 2021.

\bibitem{veaux2017superseded}
C.~Veaux, J.~Yamagishi, K.~MacDonald \emph{et~al.}, ``{Superseded-CSTR VCTK
  corpus: English multi-speaker corpus for CSTR voice cloning toolkit},'' 2017.

\bibitem{puchtler2021huiaudiocorpusgerman}
P.~Puchtler, J.~Wirth, and R.~Peinl, ``{Hui-audio-corpus-german: A high quality
  tts dataset},'' in \emph{{German Conference on Artificial Intelligence
  (K\"unstliche Intelligenz)}}, 2021.

\bibitem{muller_thorsten_2021_5525342}
T.~Müller and D.~Kreutz, ``{Thorsten - Open German Voice (Neutral) Dataset},''
  \url{https://doi.org/10.5281/zenodo.5525342}, 2021.

\bibitem{ling2021blizzard}
Z.-H. Ling, X.~Zhou, and S.~King, ``{The Blizzard Challenge 2021},'' in
  \emph{{Proc. Blizzard Challenge Workshop}}, 2021.

\bibitem{css10}
K.~Park and T.~Mulc, ``{CSS10: A Collection of Single Speaker Speech Datasets
  for 10 Languages},'' \emph{{Interspeech}}, 2019.

\bibitem{pratap2020mls}
V.~Pratap, Q.~Xu, A.~Sriram, G.~Synnaeve, and R.~Collobert, ``{MLS: A
  Large-Scale Multilingual Dataset for Speech Research},'' in
  \emph{{Interspeech}}, 2020.

\bibitem{shi2020aishell}
Y.~Shi, H.~Bu, X.~Xu, S.~Zhang, and M.~Li, ``{Aishell-3: A multi-speaker
  mandarin TTS corpus and the baselines},'' \emph{arXiv:2010.11567}, 2020.

\bibitem{infore}
``{InfoRe Technology 1},'' \url{https://github.com/TensorSpeech/TensorFlowASR},
  accessed: 2023-02-27.

\bibitem{rothauser1969ieee}
E.~Rothauser, ``{IEEE recommended practice for speech quality measurements},''
  \emph{IEEE Trans. on Audio and Electroacoustics}, 1969.

\bibitem{speechbrain}
M.~Ravanelli, T.~Parcollet, P.~Plantinga, A.~Rouhe, S.~Cornell \emph{et~al.},
  ``{SpeechBrain: A General-Purpose Speech Toolkit},'' 2021.

\bibitem{ecapa}
B.~Desplanques, J.~Thienpondt, and K.~Demuynck, ``{ECAPA-TDNN: Emphasized
  Channel Attention, Propagation and Aggregation in TDNN Based Speaker
  Verification},'' in \emph{{Interspeech}}, 2020.

\end{thebibliography}

\end{document}